\documentclass[aps,pre,twocolumn,showpacs,superscriptaddress]{revtex4}

\usepackage{graphicx}
\usepackage{latexsym}
\usepackage{amsmath}
\usepackage{amsfonts}
\usepackage{amssymb}
\usepackage{bm}
\usepackage{amsbsy}
\usepackage{color}
\usepackage{paralist}

\renewcommand{\vec}[1]{{\bf #1}}

\newcommand{\myhat}[1]{{\bf \hat{#1}}}

\def\1{{\bf \hat{I}}}

\definecolor{red}{rgb}{1,0,0}
\definecolor{green}{rgb}{0,1,0}
\definecolor{blue}{rgb}{0,0,1}

\newcommand{\first}[1]{}

\newcommand{\oneeps}[3]{\includegraphics[#1]{#2}}

\begin{document}

\title{Dynamical regimes and hydrodynamic lift of viscous vesicles under shear}
  
\author{Sebastian Me\ss linger}
\affiliation{Institut f\"ur Festk\"orperforschung, Forschungszentrum 
J\"{u}lich, D-52425 J\"{u}lich, Germany}
\author{Benjamin Schmidt}
\affiliation{Institut f\"ur Festk\"orperforschung, Forschungszentrum 
J\"{u}lich, D-52425 J\"{u}lich, Germany}
\affiliation{Division of Engineering Science, University of Toronto, 
Toronto M5S 2E4, Canada}
\author{Hiroshi Noguchi}
\affiliation{Institut f\"ur Festk\"orperforschung, Forschungszentrum 
J\"{u}lich, D-52425 J\"{u}lich, Germany}
\affiliation{Institute for Solid State Physics, University of Tokyo, 
Kashiwa, Chiba 277-8581, Japan}
\author{Gerhard Gompper}
\affiliation{Institut f\"ur Festk\"orperforschung, Forschungszentrum 
J\"{u}lich, D-52425 J\"{u}lich, Germany}
\affiliation{Institute for Advanced Simulations, Forschungszentrum 
J\"{u}lich, D-52425 J\"{u}lich, Germany}
  
\date{\today}
  
\begin{abstract}
The dynamics of two-dimensional viscous vesicles in shear flow, 
with different fluid viscosities $\eta_{\rm in}$ and $\eta_{\rm out}$ 
inside and outside, respectively, 
is studied using mesoscale simulation techniques. 
Besides the well-known tank-treading and 
tumbling motions, an oscillatory swinging motion is observed in the 
simulations for large shear rate.
The existence of this swinging motion requires the excitation of higher-order 
undulation modes (beyond elliptical deformations) in two dimensions.
Keller-Skalak theory is extended to deformable two-dimensional 
vesicles, such that a dynamical phase diagram can be
predicted for the reduced shear rate and the viscosity contrast $\eta_{\rm in}/\eta_{\rm out}$.
The simulation results are found to be in good 
agreement with the theoretical predictions, when thermal fluctuations are
incorporated in the theory.
Moreover, the hydrodynamic lift force, acting on vesicles under shear 
close to a wall, is determined from simulations for various viscosity 
contrasts. For comparison, the lift force is calculated numerically in
the absence of thermal fluctuations using the 
boundary-integral method for equal inside and outside viscosities. 
Both methods show that the dependence of the lift force on the distance 
$y_{\rm {cm}}$ of the vesicle center of mass from the wall is well 
described by an effective power law $y_{\rm {cm}}^{-2}$ for 
intermediate distances 
$0.8 R_{\rm p} \lesssim y_{\rm {cm}} \lesssim 3 R_{\rm p}$ 
with vesicle radius $R_{\rm p}$. 
The boundary-integral calculation indicates that the lift force decays 
asymptotically as $1/[y_{\rm {cm}}\ln(y_{\rm {cm}})]$ far from the wall.
\end{abstract}
 

\maketitle

\section{Introduction}

Vesicles are fluid droplets enclosed by a fluid lipid membrane. 
Typically, vesicles have sizes of the order of 100 nanometers to 
10 micrometers, whereas the 
thickness of the membrane is only of the order of a nanometer.  
Therefore, the membrane can often be regarded as a 
two-dimensional manifold. Vesicle shapes and fluctuations are then
governed by the curvature elasticity. This description has been very 
successful to explain vesicles behavior in thermal equilibrium \cite{seif97}.  
  
The dynamics of fluid vesicles in shear flow has attracted much 
attention recently    
\cite{KELLER1982,krau96,Beaucourt2004,nog04,nog05,nogu09,pozr01,haas97,kant05,Kantsler2006,
made06,Misbah2006,vlah07,Noguchi2007a,Danker2007,Lebedev2007,lebe08}.
Aspherical vesicles under shear can be found in different dynamical phases,
depending on the viscosities $\eta_{\rm {in}}$ and $\eta_{\rm {out}}$ of inner and
outer fluids, respectively, the membrane viscosity $\eta_{\rm {mb}}$, the bending
rigidity $\kappa$, the shear rate $\dot{\gamma}$, the membrane area, and 
enclosed volume. 
As long as shape relaxation times of the vesicle are small compared 
to the time scale set by the shear rate $\dot{\gamma}$, the vesicle is 
always close to its equilibrium shape.
Under these conditions, vesicles can be found either in 
tank-treading (TT) motion,
or -- if the viscosity contrast $\lambda=\eta_{\rm {in}}/\eta_{\rm {out}}$ 
exceeds a critical value --  in a tumbling (TB) motion. In the 
tank-treading regime, the vesicle shape and orientation are stationary in time, 
but the membrane rotates around the vesicle's center of mass in the 
same direction as the rotational part of the shear flow. 
Here, the orientation is characterized by the inclination angle $\theta$ with 
respect to the flow direction.  In the tumbling regime, the long axis of the 
vesicle performs a periodic rotation. 
Keller and Skalak \cite{KELLER1982} developed a theory for fluid vesicles 
with fixed ellipsoidal shape and different viscosity contrasts, which
is able to explain the observed experiments.
In recent years, computer simulations 
\cite{krau96,Beaucourt2004,nog04,nog05,nogu09} 
have shown that the Keller-Skalak (KS) theory provides indeed a very 
good description of tank-treading and tumbling.
    
However, the vesicle dynamics is far less understood when the shear rate is 
large enough that the vesicle cannot relax into its 
equilibrium shape. Only recently, it was shown that a third dynamical 
regime can appear under these conditions, the swinging (SW) regime 
\cite{Kantsler2006,made06,Misbah2006,vlah07,Noguchi2007a,Danker2007,Lebedev2007,lebe08} 
--- also called the trembling \cite{Kantsler2006} or vacillating-breathing 
regime \cite{Misbah2006}.
In the swinging state, oscillations of shape and inclination 
angle together determine the vesicle dynamics.
Swinging vesicles were first observed experimentally in 
Ref.~\cite{Kantsler2006}.  With increasing shear rate, a transition from 
tumbling to swinging motion was found.
A perturbation theory for quasi-spherical vesicles to lowest order in 
the deviation from the spherical shape predicted swinging for a range 
of viscosity contrasts \cite{Misbah2006}; however, since  
the shear rate appears only as basic (inverse) time scale in this approach, 
the experimental results could not be explained.
Therefore, higher-order expansions for quasi-spherical vesicles
\cite{Danker2007,Lebedev2007,lebe08} and a generalized 
Keller-Skalak (KS) theory for ellipsoidal vesicles \cite{Noguchi2007a} 
have been developed, which are able to 
predict phase transitions with varying shear rate and thereby to explain the 
experiments of Ref.~\cite{Kantsler2006}.

The dynamics in the TT, TB, and SW phases has been studied mainly for single 
vesicles in an unbounded fluid.
However, in particular due to its physiological importance, it is of 
high interest to study  the dynamical behavior of vesicles under shear 
in the presence of walls.  In this case, vesicles are repelled from a wall 
due to a hydrodynamic lift force $F_{\rm L}$.
The hydrodynamic lift force plays an important role in circulatory systems 
of vertebrates.  Since the lift force pushes red blood cells to the 
center of a blood vessel, where the flow velocity is largest, it increases 
the efficiency of oxygen transport.
On the other hand, white blood cells  move along the vessel walls in order
to find defects in the vascular endothelium \cite{fung97,imhof1997}. 
This is achieved by special ligands, which are located at the outside of 
white blood cells and bind to receptors 
on the vessel wall to resist the hydrodynamic lift force 
\cite{chan00,dwir03}.

The existence of a hydrodynamic lift force was first reported by Poiseuille 
in 1836 \cite{Poiseuille1836}, who observed this effect on blood cells. 
In recent years, the hydrodynamic lift force was studied intensively,
both theoretically 
\cite{Olla1997,Olla1997a,Cantat1999,Seifert1999,Sukumaran2001} 
and experimentally \cite{lorz00,Abkarian2002,abka05,call08}.
Abkarian et al.~\cite{Abkarian2002,abka05} observed the unbinding of a heavy 
vesicle, which was pulled by gravity towards a wall, with increasing 
shear rate. For vesicles which are not in direct contact
with the wall, only studies in three dimensions with equal viscosities 
of inside and outside fluids exist. Both, boundary-integral 
simulations \cite{Sukumaran2001} as well as theoretical studies
\cite{Olla1997,Olla1997a} show that the lift force decays with 
a power law $1/y^2_{\rm {cm}}$ with increasing distance between
the vesicle's center of mass and the wall. For vesicles in two dimensions, 
there are only theoretical and numerical
studies which focus on 
adhering vesicles bound to the wall by a short-ranged attractive 
potential \cite{Cantat1999,Seifert1999}.

In this paper, we study the dynamics of a two-dimensional ($2$D)
vesicle as a function of viscosity contrast $\lambda$ and shear rate 
$\dot\gamma$, both in the bulk and near a wall. 
The advantage of simulations of a vesicle in two dimensions is (i) 
the reduced numerical effort of hydrodynamics simulations, which allows 
for larger system sizes, longer accessible time scales, and better 
statistics, and (ii) the simpler form of the equations of the KS theory, 
where no integrals remain in the geometric factors -- 
unlike in the $3$D version (see App.~\ref{sec:KS_derivation}). This 
facilitates a detailed comparison of the results of theory and 
simulations.
Using a mesoscopic hydrodynamics approach, 
we first show that the SW mode also exists in two dimensions, and determine the 
dynamical phase diagram. The simulation results are compared with the 
predictions of a generalized KS theory.
Second, we study the lift force $F_{\rm L}$ of $2$D vesicles, by covering 
the full range of wall distances $y_{\rm {cm}}$, and investigate the 
effects of viscosity contrast $\lambda$. 
Moreover, we investigate the effect of a wall on the TT-TB
behavior. For comparison with the results of mesoscopic 
hydrodynamics simulations, 
we also determine $F_{\rm L}$ and the inclination angle $\theta$ by the 
boundary-integral method for tank-treading vesicles with $\lambda=1$.

\section{Theory and Methods}    

\subsection{Dimensionless Parameters}

In a $2$D vesicle, the perimeter $L_{\rm p}$ and the enclosed area
$A$ are kept constant
(analogously to the constant membrane surface and the enclosed volume of
$3$D vesicles). It is useful to
combine these two parameters into a dimensionless quantity, the reduced area
\begin{equation} \label{equ:reducedarea}
  A^*:=\frac{4\pi A}{{L_{\rm p}}^2}= \Big(\frac{R_{\rm A}}{R_{\rm p}}\Big)^2.
\end{equation}
Here $R_{\rm p}=L_{\rm p}/2\pi$ and $R_{\rm A}=\sqrt{A/\pi}$ are the radii of 
circles with the same $L_{\rm p}$ and $A$ as those of the vesicle, respectively. 
$A^*$ is the ratio between the enclosed area $A$ and the area of a circle 
with the same perimeter $L_{\rm p}$.  We focus here on 
a reduced area of $A^*=0.7$ as a representative for vesicles which
deviate significantly from the circular shape.  
   
Shape and orientation of the vesicles
are quantified by a shape parameter $\alpha$ and inclination angle $\theta$ 
based on the gyration tensor of the vesicle membrane. When
$\Lambda_{\rm {max}}$ and $\Lambda_{\rm {min}}$  are the two eigenvalues of the 
gyration tensor ($\Lambda_{\rm {max}}\ge \Lambda_{\rm {min}}$), and 
$\myhat{e}_{\rm {max}}$ and $\myhat{e}_{\rm {min}}$ the corresponding 
eigenvectors, the ``asphericity" is described by 
$\alpha=(\Lambda_{\rm {max}}-\Lambda_{\rm {min}})/(\Lambda_{\rm {max}}+\Lambda_{\rm {min}})$
and the vesicle orientation by the inclination angle 
$\theta=\measuredangle(\myhat{x},\myhat{e}_{\rm {max}})$, where $\myhat{x}$ 
is the shear and $\myhat{y}$ the gradient direction.
    
The stability of dynamical phases mainly depends on two parameters, 
the viscosity contrast $\lambda$ and the reduced shear rate
\begin{equation} \label{equ:redshear}
  \dot{\gamma}^*:=\frac{\dot{\gamma}\eta_{\rm {out}} R_{\rm p}^{3}}{\kappa}.
\end{equation}
The time $\eta_{\rm {out}} R_{\rm p}^{3}/\kappa$
is the characteristic relaxation time in thermal equilibrium,
where $\kappa$ is the bending rigidity. Thus
$\dot{\gamma}^*$ expresses the interplay between the perturbation by the 
external field $\dot{\gamma}$ and the ability of the vesicle to
restore its equilibrium shape.

\subsection{Generalized Keller-Skalak Theory in Two Dimensions}
  
Keller-Skalak (KS) theory \cite{KELLER1982} is based on the 
assumption that vesicles have a fixed ellipsoidal shape. Therefore, it  
cannot describe the swinging state with oscillating vesicle shapes. 
Therefore, KS theory has been generalized
to include shape deformation in three dimensions \cite{Noguchi2007a}. 
This theory is applicable to ellipsoidal vesicles 
over a wide range of reduced volumes, while higher-order perturbation theory 
\cite{Danker2007,Lebedev2007,lebe08} is limited to quasi-spherical vesicles.
Here, we employ the two-dimensional version of the 
generalized KS theory.  The differential equations for the
asphericity $\alpha$ and inclination angle $\theta$ are given by
\begin{eqnarray}
  \frac{1}{\dot{\gamma}}\frac{d\alpha}{dt} & = & 
     -\frac{b_0}{A^*\dot{\gamma}^*} \frac{R_{\rm p}}{\kappa}
                 \frac{\partial F}{\partial \alpha} + b_1\sin (2\theta),
       \label{equ:alphaevolution}\\
  \frac{1}{\dot{\gamma}}\frac{d\theta}{dt} & = & 
                 \frac{1}{2}\left[-1+B(\alpha) \cos(2\theta)\right],
      \label{equ:thetaevolution}
\end{eqnarray}
with prefactors 
\begin{equation} \label{eq:b_factors}
b_0=\frac{3}{4\pi(\lambda+1)} \ \ \ \mathrm{and} \ \ \ b_1=\frac{3}{2(\lambda+1)}.
\end{equation}
There are no adjustable parameters.
An explicit expression for  $B(\alpha)$ and its derivation are described 
in App.~\ref{sec:KS_derivation}.
  
The time evolution of $\theta$ is described by Eq.~(\ref{equ:thetaevolution}),
which has the same form as in two-dimensional KS theory.
However, $B(\alpha)$ is now not constant but depends on the time-dependent 
vesicle shape $\alpha(t)$.  The time evolution of $\alpha$ 
(see Eq.~(\ref{equ:alphaevolution})) is derived based on the perturbation
theory of quasi-circular vesicles~\cite{fink08}.
Here, $F$ is the free energy of the vesicle shape at constant $A^*$.
$F$ attains its minimum for an elliptical vesicle shape in equilibrium. 
Thus, the first term on the right hand side of Eq.~(\ref{equ:alphaevolution}) 
causes a relaxation of $\alpha$ towards its equilibrium value. The second 
term represents the change of $\alpha$ due to the external flow field. 
Eqs.~(\ref{equ:alphaevolution}) and (\ref{equ:thetaevolution})
are solved numerically using a fourth-order Runge-Kutta method. 
  
Thermal fluctuations can be incorporated in this approach by adding 
Gaussian white noises $g_{\alpha}(t)$ and $g_\theta(t)$ to 
Eqs.~(\ref{equ:alphaevolution}) and (\ref{equ:thetaevolution}), respectively. 
The noise terms obey the
fluctuation-dissipation theorem, such that $\left<g_i(t)\right>=0$ and 
$\left<g_i(t)g_j(t')\right>=(2k_{\rm B}T/\zeta_i)\delta_{i,j}\delta(t-t')$  
with $i,j \in \{\alpha,\theta\}$, where $k_{\rm B}T$ is the thermal energy.
As a reasonable approximation, we employ the rotational
friction coefficients of a circle, 
\begin{equation}
  \zeta_\alpha= \frac{4\pi}{3} \eta_{\rm {out}} {R_{\rm A}}^2(\lambda+1)
     \ \ \mathrm{and} \ \ 
  \zeta_\theta= 4\pi \eta_{\rm {out}} {R_{\rm p}}^2.
\end{equation}

\subsection{Mesoscale Hydrodynamics Simulation Method}
\label{sec:vessimdetails}
\subsubsection{Membrane Model}

The membrane is modeled by a closed chain of $n$ monomers of mass $M$. 
For a monomer with index $i$ (with $1\le i\le n$), we introduce the notation 
\begin{equation}
  i_{-}=(i-1)\,\mathrm{mod}\, n \qquad \mathrm{and} \qquad 
  i_{+}=(i+1)\,\mathrm{mod}\, n
\end{equation}
for the indices of its two neighboring monomers.
Thereby, the ring topology is taken into account correctly. The monomers 
are connected by a harmonic spring potential
\begin{equation}
\label{equ:bond_pot_ves}
  U_{\rm {sp}}=\frac{k_{\rm {sp}}}{2}\sum^{n}_{i=1}(|\vec{R}_i|-l)^2 , 
\end{equation}
where 
$\vec{R}_i:=\vec{r}_{i_{+}}-\vec{r}_i$
are the bond vectors, and $l$ is the relaxed bond length.
The curvature elasticity of the membrane is described by the bending potential
\begin{equation} \label{equ:bendind_pot_ves}
  U_{\rm {bend}}=\frac{\kappa}{l}\sum^{n}_{i=1}
     \left(1-\frac{\vec{R}_{i_+}\cdot\vec{R}_i}{|\vec{R}_{i_+}\parallel\vec{R}_i|}\right).
\end{equation}
An area potential  
\begin{equation}
  U_{\rm A}=\frac{k_{\rm A}}{2}\left(A-A_0\right)^2.\label{equ:areapot}
\end{equation}
is introduced to control the deviations of the area $A$ from its target 
value $A_0$. 
Here, the enclosed area $A$ in Eq.~(\ref{equ:areapot}) is obtained from the
monomer positions by
\begin{equation}
  A=\frac{1}{2}\myhat{z}\cdot \sum^n_{i=1}\vec{r}_i\times\vec{r}_{i_{+}}.
\end{equation}
      
\subsubsection{Multi-Particle Collision Dynamics}

For the solvent hydrodynamics, we employ multi-particle collision 
dynamics (MPC), a particle-based mesoscopic simulation technique
\cite{kap99,kapr08,gg:gomp09a}.
The dynamics of an MPC fluid evolves in two alternating
steps. In the ``streaming step'', particles move ballistically for a 
time $\Delta t$, the collision time, according to their current velocities. 
For the ``collision step'', solvent particles are first sorted
into the cells of linear size $a$ of a regular square 
lattice; all particles in a cell then exchange momenta such that the 
total translational momentum is conserved in each collision cell.

Several modifications of the original MPC algorithm have been introduced 
recently \cite{Noguchi2007}, which differ in the way the collision step
is executed. We employ the MPC-AT$+a$ version of multi-particle
collision dynamics, which uses an Anderson thermostat (AT) and 
locally conserves angular momentum ($+a$) in addition to 
translational momentum. In MPC-AT, new particle velocities relative to 
the center-of-mass velocity are chosen  
from a Maxwell-Boltzmann distribution with temperature $T$. This
thermostat avoids any heating due to energy dissipation in sheared system. 
For details of the MPC-AT$+a$ algorithm, see Refs.~\cite{Noguchi2007,nogu08}. 
We use this algorithm, since
local angular-momentum conservation is crucial in binary fluid systems 
with different viscosities \cite{Gotze2007}.
      
Simulations are performed with a rectangular simulation box with linear sizes 
$L_x$ and $L_y$, periodic boundary conditions in the
$x$ direction, and no-slip wall boundary conditions in the $y$ direction.
Linear shear flow with shear rate $\dot{\gamma}$ is realized by moving 
the upper wall with a velocity $\dot{\gamma}L_y\myhat{x}$, whereas the 
lower wall is held at rest.
      
Many properties of the MPC-AT$+a$ solvent can be adjusted by the simulation 
parameters collision time $\Delta t$, the particle number density $n_{\rm s}$, 
and the particle mass $m$. 
The solvent viscosity $\eta=\eta_{\rm {kin}}+\eta_{\rm {coll}}$ is a sum 
of a kinetic $\eta_{\rm {kin}}$ and a collisional
contribution $\eta_{\rm {coll}}$, which have been calculated analytically 
\cite{nogu08},
\begin{eqnarray} \label{equ:MPCATa_analytic_eta1}
  \eta_{\rm {kin}} &=& \frac{n_{\rm s} k_{\rm B}T \Delta t}{ a^2 }
      \left[ \frac{n_{\rm s}}{n_{\rm s}- 1 } - \frac{1}{2} \right], \\
       \eta_{\rm {col}} &=& \frac{m(n_{\rm s}-7/5)}{24 \Delta t}.
   \label{equ:MPCATa_analytic_eta2}
\end{eqnarray}
The viscosity $\eta_{\rm {out}}$ of the fluid outside of the vesicle is 
adjusted by varying the collision time $\Delta t$ in the range from
$\Delta t=0.003a\sqrt{m/k_{\rm B} T}$ to $\Delta t=0.01a\sqrt{m/k_{\rm B} T}$.
Since for these collision times the mean free path is much smaller than
the cell size $a$,  the total shear viscosity $\eta$ is dominated 
by $\eta_{\rm {coll}}$  (see Eqs.~(\ref{equ:MPCATa_analytic_eta1}) and
(\ref{equ:MPCATa_analytic_eta2})).
Since the collisional viscosity $\eta_{\rm {coll}} \propto m$,
the viscosity contrast $\lambda$ can be varied by using different masses 
$m_{\rm {in}}$ and $m$ of the inner and outer fluid particles, respectively,
which implies
\begin{equation}
  \lambda=\frac{\eta_{\rm {in}}} {\eta_{\rm {out}}} 
                     \approx \frac{m_{\rm {in}}}{m}.
\end{equation}
In our simulations, the viscosity contrast is varied from $\lambda=1$ to 
$\lambda=10$ (with $m\le m_{\rm {in}}\le 10 m$), while all the other 
MPC parameters are the same for the fluid on both sides of the membrane.

\subsubsection{Membrane Interactions}
\label{sec:memb_int}

In order to describe an impermeable membrane in flow, 
it has to be ensured that MPC particles stay on the correct side of 
the membrane ({\em i.e.} inside or outside of the vesicle). 
For numerical efficiency, it is advantageous to relax this condition for 
short length and time scales, as it was done in 
previous $3$D vesicle simulations \cite{nog05}. 
The streaming and collision steps for the fluid particles are carried 
as in the absence of the membrane. This implies that after each 
streaming step, some MPC particles have crossed the membrane. 
For the (few) particles which are now located on the wrong side of the 
membrane, with a direction of their velocity which would bring them away 
even further away from the membrane, the velocities have to be modified 
such that they move towards the membrane instead, in order to cross 
back to their correct side. We denote this velocity update a 
``membrane collision''.  It has to be constructed such that the 
translational and angular momentum as well as the kinetic energy 
of the fluid particles and membrane monomers are conserved locally. 
Our procedure for membrane collisions is a generalization of the 
standard bounce-back rule for no-slip boundary conditions.
A detailed description of this procedure is provided in 
App.~\ref{sec:membrane_coll}.

In order to prevent the membrane from crossing the walls, a purely 
repulsive Lennard-Jones potential  
\begin{equation}
  U_{\rm w}(y)\!=\!\left\{\!
  \begin{array}{ll}
    \setlength{\tabcolsep}{-1cm}
    4\varepsilon \left[ \left(\frac{\sigma}{y}\right)^{12} - 
           \left(\frac{\sigma}{y}\right)^6\right]+\varepsilon, 
             & 0 \le y \le \sqrt[6]{2}\sigma\\
          0, & \mathrm{otherwise}\\
  \end{array}
  \right.\nonumber
\end{equation}
is employed, which depends only on the distance $y$ of a monomer 
from a wall.
      
For the determination of hydrodynamic lift forces, we employ a 
gravitational body force $\vec{f}_{\rm G}=-\myhat{y}g\Delta\varrho$,
which acts on the internal fluid of the vesicle. Here, $g$ denotes the 
strength of the gravitational field, and $\Delta\varrho$ is the mass-density 
difference between the inner and outer fluids. 
The gravitational body force $\vec{f}_{\rm G}$ acting on the inner fluid
can be expressed as a potential $U_{\rm G}$, which only depends on the 
monomer positions,
\begin{equation} \label{equ:gravpot}
  U_{\rm {G}}=\frac{F_{\rm G}}{6A}\sum_i\left(y_i+y_{i_+}\right) 
      \left(\vec{r}_i\times\vec{r}_{i_+}\right)\cdot\myhat{z}.
\end{equation}
Here, $y_i$ and $y_{i_+}$ are the $y$ components of the monomer positions 
$\vec{r}_{i}$ and $\vec{r}_{i_+}$, respectively,
and $F_{\rm G}=\left|\int_A \vec{f}_{\rm G} dA \right|$ is the total 
gravitational force acting on the vesicle. $F_{\rm G}$ has a constant 
value and is used as a simulation parameter.
  
As long as not specified otherwise, the parameters used in our vesicle 
simulations are $n=50$, $l=a=\sqrt[6]{2}\sigma$, $n_{\rm s}=10a^{-2}$, 
$M=10m$, $\varepsilon=10k_{\rm B}T$, and $\kappa/l=50 k_{\rm B}T$. For 
the reduced area, we require that it deviates less than $1\%$ from its 
target value of $A^*=0.7$. Since $A^*$ is a function of the perimeter 
$L_{\rm p}$ and the enclosed area $A$ (see Eq.~(\ref{equ:reducedarea})), 
the parameters $k_{\rm {sp}}$ and $k_{\rm A}$ 
for the potentials $U_{\rm {sp}}$ and $U_{\rm A}$, respectively, have to be 
sufficiently large. We chose
$k_{\rm {sp}}=10^4 k_{\rm B}T/a^2$ and $k_{\rm A}=80 k_{\rm B}T/a^4$.
With these parameters, the effective vesicle radius is obtained to be
$R_{\rm p}=7.8\,l$.
The size of the simulation box is $L_x=L_y=80a$.  Gravitational 
forces $F_{\rm G}$ are only applied in simulations for the hydrodynamic 
lift force, where values in the range
$k_{\rm B}T/a\le F_{\rm G} \le 50 k_{\rm B}T/a$ are investigated. 
  
In simulations, different reduced shear rates $\dot{\gamma}^*$ can 
be achieved, according to Eq.~(\ref{equ:redshear}), by varying 
$\dot{\gamma}$, $\eta_{\rm {out}}$, $R_{\rm p}$, or $\kappa$. 
Since equilibrium properties like the undulation spectrum depend on
$R_{\rm p}$ and $\kappa$, we 
vary $\dot{\gamma}^*$ by adjusting $\dot{\gamma}$ and 
$\eta_{\rm {out}}$. 
In order to avoid inertial effects, we restrict the shear rates to
obtain low Reynolds numbers Re\ $=\dot\gamma\rho R_{\rm p}^2/\eta_{\rm {out}}$,
where $\rho$ is the density of the outer fluid.
The maximum Reynolds number is Re$=0.17$.

\subsection{Boundary-Integral Method}
\label{sec:oseen}

For comparison with our MPC simulation results 
of the lift force, we also perform numerical boundary-integral calculations.  
The hydrodynamic lift force in $2$D 
has been studied previously with the boundary-integral approach for vesicles 
in direct contact with the wall \cite{Cantat1999,Seifert1999}. 
This method has the advantage that it can be used to 
calculate lift forces on vesicles even for very large distances
$y_{\rm {cm}}$ from the wall and for reduced areas $A^*$ close to 
unity, which are not easily accessible by MPC simulations. 
On the other hand, our boundary-integral calculation is restricted 
to elliptical shapes and ignores thermal fluctuations, which give rise,
{\em e.g.}, to undulation-induced repulsion near a wall. 
We focus on tank-treading elliptical vesicles without  
viscosity contrast, {\em i.e.} $\lambda=1$. 
In the steady tank-treading state the lift force can calculated from a 
single, time-independent vesicle shape.
Whereas in MPC simulations, the wall distance $y_{\rm {cm}}$ is 
calculated for a given strength of the gravitational force, 
we follow the opposite procedure with the boundary-integral approach,
by calculating the lift force for a given wall distance $y_{\rm {cm}}$.

For ellipse half axes $a_1$ and $a_2$, wall distance
$y_{\rm {cm}}$, and inclination angle $\theta$, the location ${\bf r}$ 
of the vesicle membrane is uniquely defined (with $x_{\rm {cm}}\equiv 0$).
With a parameterization by the angle $\varphi$, it is
\begin{equation}
  \vec{r}(\varphi)=y_{\rm {cm}}\myhat{y}+\left(
            \begin{array}{cc}
               \cos\theta & -\sin\theta\\
               \sin\theta & \cos\theta
            \end{array}
          \right)\vec{r}'(\varphi)
\end{equation}
where $\vec{r}'(\varphi)$ is the membrane position in the principal-axis 
system of the ellipse,
\begin{equation}
  \vec{r}'(\varphi)=\left(
            \begin{array}{c}
                a_1 \cos\varphi\\
                a_2 \sin\varphi
            \end{array}
          \right).
\end{equation}
  
\begin{figure}[t]
   \oneeps{width=7.8cm}{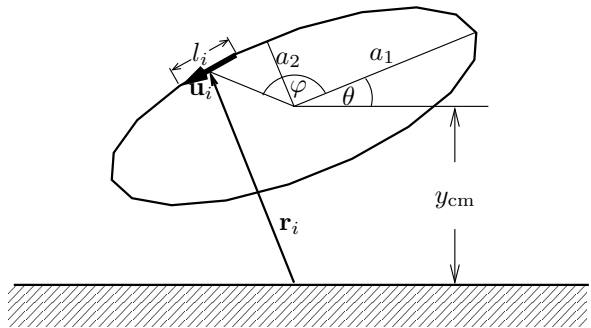}{
    }
\caption{For the boundary-integral calculation, the vesicle is discretized 
into segments $i$ of length $l_i$, which have the orientation 
$\myhat{u}_i$ and a center-of-mass position $\vec{r}_i$.}
\label{fig:oseen_drawing}
\end{figure}
  
In the steady tank-treading state, the center-of-mass velocity
$\vec{v}_{\rm {cm}}=\myhat{x}v_{\rm {cm}}$ of the vesicle has only a
non-vanishing component in shear direction. If $v_{\rm {cm}}$ and the 
tank-treading angular velocity $\omega$ are known, the velocity of
the tank-treading membrane is given by 
\begin{equation} \label{eq:tt_velocity}
  \vec{v}(\varphi)=\myhat{x}v_{\rm {cm}}+{R_{\rm p}\omega}\myhat{t}(\varphi).
\end{equation}
with the tangent vector
\begin{equation} 
   \myhat{t}(\varphi)=\left(
     \begin{array}{cc}
         \frac{a_2}{a_1}\sin\theta & \frac{a_1}{a_2}\cos\theta\\
        -\frac{a_2}{a_1}\cos\theta & \frac{a_1}{a_2}\sin\theta
     \end{array}
     \right) \frac{\vec{r}'(\varphi)}{\left|\vec{r}'(\varphi)\right|}.
\end{equation}
In a tank-treading membrane in shear flow, forces arising from pressure and
viscous stress have to be balanced in order to maintain a steady motion. 
The force distribution $\vec{f}(\vec{r}')$ along the membrane $\partial A$ 
is related to the velocity field at position $\vec{r}$ by 
\begin{equation} \label{eq:intequ}
   \vec{v}(\vec{r})-\dot{\gamma}y\myhat{x}=\int_{\partial A}\mathbf{G}(\vec{r},\vec{r}'(s))\vec{f}(\vec{r}'(s))ds.
\end{equation}
Here, $ds$ is a line element of the membrane $\partial A$, and
the second-order tensor $\mathbf{G}(\vec{r},\vec{r}'(s))$ is the
Greens function of the Stokes equation which satisfies the boundary conditions. 
For vesicles in an unbounded fluid, $\mathbf{G}(\vec{r},\vec{r}'(s))$ is 
the Oseen tensor.  
In our case of a vesicle near a wall, the half-space Oseen tensor -- 
also known as Blake tensor -- is convenient, as it realizes no-slip 
boundary conditions at the wall. The full expression for the two-dimensional 
Blake tensor can be found in Ref.~\cite{Pozrikidis1992}.

The difficulty is that the force distribution $\vec{f}(\vec{r}')$ along 
the membrane is a priori unknown. Instead, we know the velocities 
$\vec{v}(\vec{r})$ at each site of the membrane. 
Eq.~(\ref{eq:intequ}) is thereby a Fredholm integral equation of the 
first type.  This integral equation is solved numerically. For this 
purpose, we discretize the membrane in $N$ straight segments, which have 
to be small enough such that the difference in velocities 
between two neighboring segments is small and the force distribution 
can be assumed to be constant along the segment.
A segment with index $i$ has a velocity $\vec{v}_i$, center-of-mass 
position $\vec{r}_i$, length $l_i$ and orientation $\myhat{u}_i$ 
(see Fig.~\ref{fig:oseen_drawing}).
The discretized form of Eq.~(\ref{eq:intequ}) is
\begin{equation}
\vec{v}_i-\dot{\gamma}y_i\myhat{x}=
  \sum_{j=1}^N\int_{-l_j/2}^{l_j/2}\mathbf{G}(\vec{r}_i,\vec{r}_j+\myhat{u}_j s)
          \vec{f}(\vec{r}_j+\myhat{u}_j s)ds.
\end{equation}
Since the force distribution $\vec{f}_j=\vec{f}(\vec{r}_j+\myhat{u}_j s)$ is 
assumed to be constant over the whole segment $j$, it can be moved outside of 
the integral,
\begin{equation}
\vec{v}_i-\dot{\gamma}y_i\myhat{x}=
  \sum_{j=1}^N \left[\int_{-l_j/2}^{l_j/2}
      \mathbf{G}(\vec{r}_i,\vec{r}_j+\myhat{u}_j s)ds\right] \vec{f}_j 
    =\sum_{j=1}^N \mathbf{H}_{ij} \vec{f}_j.
\end{equation}
The calculation of $\mathbf{H}_{ij}$ can be performed analytically,  
both for the free-space and the half-space Oseen tensor.
Thus, the integral equation~(\ref{eq:intequ}) is reduced to a set of
linear algebraic equations which can be easily solved numerically. 
     
The segment velocities $\vec{v}_i$ depend linearly on $\omega$ and
$v_{\rm {cm}}$ (see Eq.~(\ref{eq:tt_velocity})). Therefore, we can extend
the linear system of equations~(\ref{eq:intequ}) by two additional 
conditions, which determine  
$\omega$ and $v_{\rm {cm}}$ self-consistently in the steady state. 
For the first condition,
we require that the sum of tangential forces along the membrane vanishes.
The second condition is that the vesicle does not experience a net force 
in shear direction. 
The total system of linear equations finally reads
\begin{eqnarray}
  -\dot{\gamma}y_i\myhat{x}&=&\sum_{j=1}^N \mathbf{H}_{ij}\vec{f}_j
             - v_{cm}\myhat{x}-\myhat{t}_i R_{\rm p}\omega\\
       0 & = & \sum_{i=1}^N \myhat{x}\cdot\vec{f}_i l_i\\
       0 & = & \sum_{i=1}^N\myhat{t}_i\cdot \vec{f}_i l_i.
\end{eqnarray}
This set of equations is solved numerically with up to $N=600$ segments. 
Once, $\omega$, $v_{\rm {cm}}$ and the force distribution are known, 
quantities like the lift force $\vec{F}_{\rm L}$ and the torque $\vec{M}$ 
on the vesicle, as well as the velocity $\vec{v}(\vec{r})$
and pressure fields $p(\vec{r})$ in the surrounding fluid can be calculated 
as 
\begin{eqnarray}
  \vec{F}_{\rm L}&=&\sum_{j=1}^N\vec{f}_j l_j,
             \label{eq:lift_force}\\
  \vec{M}&=&\sum_{j=1}^N\left(\vec{r}_j-y_{\rm {cm}}\myhat{y} \right)
        \times \vec{f}_j l_j, \label{eq:torque}\\
  \vec{v}(\vec{r})&=&\dot{\gamma}y\myhat{x} 
       + \sum_{j=1}^N \left[\int_{-l_j/2}^{l_j/2}
       \mathbf{G}(\vec{r},\vec{r}_j+\myhat{u}_j s)ds\right]\vec{f}_j,\\
  p(\vec{r})&=&\sum_{j=1}^N \left[\int_{-l_j/2}^{l_j/2}
         \vec{g}(\vec{r},\vec{r}_j+\myhat{u}_j s)ds\right]\cdot\vec{f}_j.
\end{eqnarray}
Here, $\vec{g}(\vec{r},\vec{r}')$ is the half-space pressure vector 
(see Ref.~\cite{Pozrikidis1992}).
The lift force $\vec{F}_{\rm L}$ and the torque $\vec{M}$ 
(see Eqs.~(\ref{eq:lift_force}) and (\ref{eq:torque})), are thereby
functions of the four parameters $\theta$, $y_{\rm {cm}}$, $a_1$, and $a_2$, 
which define the membrane location uniquely. Using a numerical root finder 
(Brent's method), the stable inclination angle $\theta$, for which the torque 
vanishes, is determined while keeping the other parameters 
$y_{\rm {cm}}$, $a_1$, and $a_2$ fixed.

\section{Dynamical Regimes of Viscous Vesicles in Unbounded Shear Flow}

\subsection{Phase Diagram}

\begin{figure}
  \oneeps{width=8.6cm}{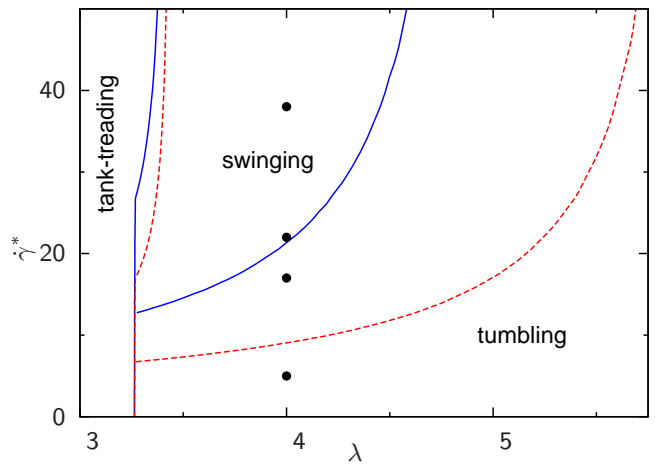}{
   }
\caption{
(Color online) Phase diagram at $A^*=0.7$. 
Dashed (red) and solid (blue) lines represent the results of the
generalized Keller-Skalak theory with $b_1=3/[2(\lambda+1)]$
(see Eq.~(\ref{eq:b_factors})) and $b_1=1/(\lambda+1)$, respectively.
The black circles ($\bullet$) indicate the location
of the simulation which are shown in Fig.~\ref{fig:dynregsimtheo}.
\label{fig:hirochisphasediagr}
}
\end{figure}

We consider first the dynamics of vesicles in shear flow, far from walls
and in the absence of a gravitational field.
The $2$D generalized KS theory predicts a phase diagram, see
Fig.~\ref{fig:hirochisphasediagr}, 
which shows the qualitatively the same features as a function of 
$\dot{\gamma}^*$ and $\lambda$ as the $3$D version~\cite{Noguchi2007a}.
At small and large $\lambda$, a vesicle exhibits tank-treading (TT) and
tumbling (TB) motion, respectively.
At large $\dot{\gamma}^*$ and intermediate $\lambda$, the swinging (SW) 
phase appears.  As in $3$D generalized KS theory,
TT with negative inclination angles $\theta<0$ appears close to the 
TT-SW transition line.  The coexistence of two TT states (one with 
$\theta<0$ and the other with $\theta>0$) or of a TT and a SW states 
are also seen.  In $2$D, TT with $\theta<0$ is stable, unlike in 
$3$D \cite{Noguchi2007a}, where the vesicle can escape by turning its
longest axis into the vorticity direction.

\begin{figure}
   \oneeps{width=8.6cm}{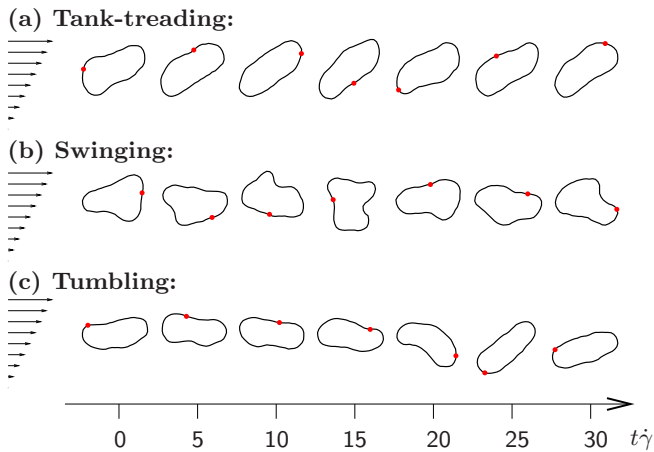}{}
\caption{(Color online) 
Sequences of vesicle snapshots for each of the dynamical regimes, 
shown (a: TT, b: SW, c: TB). A (red) bullet marks one fixed membrane element 
to indicate the membrane motion. All systems share the parameters 
$\kappa/l=50k_{\rm B}T$ and $A^*=0.7$.
Further parameters are
(a) $\eta_{\rm {out}}=36\sqrt{k_{\rm B} T m}/a$, $\lambda=1$,
$\dot{\gamma}=0.01\sqrt{k_{\rm B} T/m}/a$ corresponding to $\dot{\gamma}^*=3.6$; 
(b) $\eta_{\rm {out}}=120\sqrt{k_{\rm B} T m}/a$, $\lambda=4$, 
$\dot{\gamma}=0.0333\sqrt{k_{\rm B} T/m}/a$ corresponding to 
$\dot{\gamma}^*=38$; and
(c) $\eta_{\rm {out}}=36\sqrt{k_{\rm B} T m}/a$, $\lambda=10$, 
$\dot{\gamma}=0.01\sqrt{k_{\rm B} T/m}/a$ corresponding to 
$\dot{\gamma}^*=3.6$.
}
\label{fig:vessnapshots}
\end{figure}

MPC simulation results of the three dynamical regimes (TT, TB, SW) are 
illustrated by a sequence of snapshots in Fig.~\ref{fig:vessnapshots}. 
Fig.~\ref{fig:vessnapshots}(a) shows a tank-treading vesicle, which has 
a constant shape and orientation (except for its thermal membrane 
undulations).  However, as a marker on the membrane indicates, the 
membrane rotates around the center of mass. A tumbling vesicle is shown in 
Fig.~\ref{fig:vessnapshots}(b), where the shape remains almost unchanged, 
but the orientation steadily rotates. The motion of a marker on the 
membrane shows that the membrane is not completely fixed with respect 
to the vesicle shape. Fig.~\ref{fig:vessnapshots}(c) illustrates the 
swinging state (see also movie \cite{movie_S1}). 
As long as the vesicle orientation has a positive inclination angle, the 
elongational part of the shear flow causes an elongation of the vesicle shape
(increasing $\alpha$). For a large shape parameter $\alpha$, the vesicle 
is temporarily in the tumbling regime, until a negative inclination 
angle $\theta$ is reached (for $0\le t\dot{\gamma}\le 10$
in Fig.~\ref{fig:vessnapshots}(c)). For negative $\theta$, the elongational 
component of the flow acts to reduce $\alpha$. Due to the constraint of 
fixed perimeter $L_{\rm p}$ and fixed enclosed area $A$, the vesicle assumes a 
potato-like shape, 
such that $\alpha$ decreases (for $10\le t\dot{\gamma}\le 15$ in 
Fig.~\ref{fig:vessnapshots}(c)). The vesicle is then stretched 
again by the elongational flow leading to a positive inclination angle 
$\theta$ and increasing shape parameters $\alpha$  
(for $t\dot{\gamma} \gtrsim 20$).

It is important to note that elliptical deformations are not sufficient 
in $2$D, because the constraints on perimeter 
and area complete determine the elliptical shape \cite{fink08} --- in 
contrast to $3$D, where the deformation in the vorticity direction
provides sufficient degrees of freedom \cite{Noguchi2007a}. 
Thus, higher-order undulation modes beyond elliptical deformation 
are required, which can be seen clearly in Fig.~\ref{fig:vessnapshots}(c). 
This is reminiscent of the behavior of $3$D vesicles in an elongational
flow after flow reversal \cite{kant07,turi08}, where also higher-order
undulation modes play an important role.

\subsection{TT-TB Transition}

The generalized KS theory predicts that for small shear rates,
with $\dot{\gamma}^* \lesssim 6$, the TT-TB transition at
$\lambda \simeq 3.25$ hardly depends on $\dot{\gamma}^*$ 
(see Fig.~\ref{fig:hirochisphasediagr}). 
In this regime, shape deformations are very small, and 
the behavior can be well described by the original KS theory. 
We choose a shear rate $\dot{\gamma}=0.01\sqrt{k_{\rm B} T/m a^2}$
in our simulation,
corresponding to a small reduced shear rate $\dot{\gamma}^*=3.6$.  

\begin{figure}
    \oneeps{width=8.6cm}{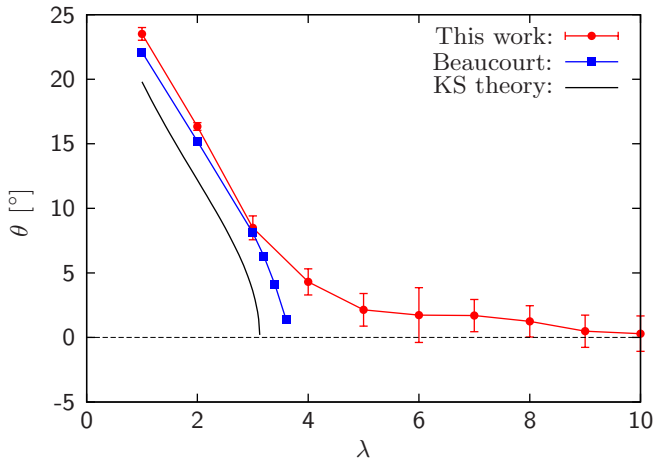}{
   }
\caption{(Color online)
Inclination angle $\theta$ as a function of viscosity contrast $\lambda$ 
for simulations with $\dot{\gamma}^*=3.6$. For comparison the results 
of the boundary-integral calculation of 
Beaucourt et al.~\cite{Beaucourt2004} (without thermal fluctuations) 
as well as the curve of KS theory \cite{KELLER1982} 
(see Eq.~(\ref{equ:stationarytheta})) are shown.
}
\label{fig:KScurve}
\end{figure}

In Fig.~\ref{fig:KScurve}, the dependence of the average inclination angle 
$\theta$ on the viscosity contrast $\lambda$ is shown. Our 
MPC simulations well reproduce the results of previous boundary-integral
calculations by Beaucourt et al.~\cite{Beaucourt2004}.
Deviations close to the TT-TB transition at $\lambda^*\simeq 4$ 
arise from thermal membrane undulations, which are present 
in our simulations, 
whereas the results of Ref.~\cite{Beaucourt2004} have been obtained in the 
zero-temperature limit.

Moreover, thermal fluctuations lead to a continuous crossover rather than a 
sharp TT-TB  transition. Thus, there are a few tumbling events already for 
viscosity contrast $\lambda=3$, and also simulations with 
$\lambda>\lambda^*\simeq 4$ exhibit some time intervals of tank-treading 
motion. Our simulations also show that the existence of a tumbling regime 
depends sensitively on the Reynolds number Re. For Re\ $\gtrsim 1$, 
$\theta$ decreases more gradually with increasing $\lambda$, 
and no tumbling motion was observed at viscosity contrasts as large as 
$\lambda=10$. Thus, we conclude that inertial effects enhance the 
TT-membrane rotation.

Fig.~\ref{fig:KScurve} also shows that KS theory \cite{KELLER1982} 
provides a good description of the $\lambda$ dependence of $\theta$ 
and the TT-TB transition. This transition is explained by the  KS theory 
as follows.  The stationary inclination angle $\theta$ in the 
tank-treading regime is determined by Eq.~(\ref{equ:thetaevolution}) 
with fixed $\alpha$ as 
\begin{equation}
\theta=-\frac{1}{2}\arccos\left(-\frac{1}{B}\right).\label{equ:stationarytheta}
\end{equation}
For small $\lambda$, the inclination angle $\theta$ decreases 
monotonically up to a critical viscosity contrast $\lambda^*$, where 
$\theta=0$. For larger viscosity contrasts $\lambda>\lambda^*$, 
the tumbling regime, there is 
no real solution of Eq.~(\ref{equ:stationarytheta}),
{\em i.e.} no stationary inclination angle exists, and the vesicle
permanently rotates.

\begin{figure}
\oneeps{width=8.6cm}{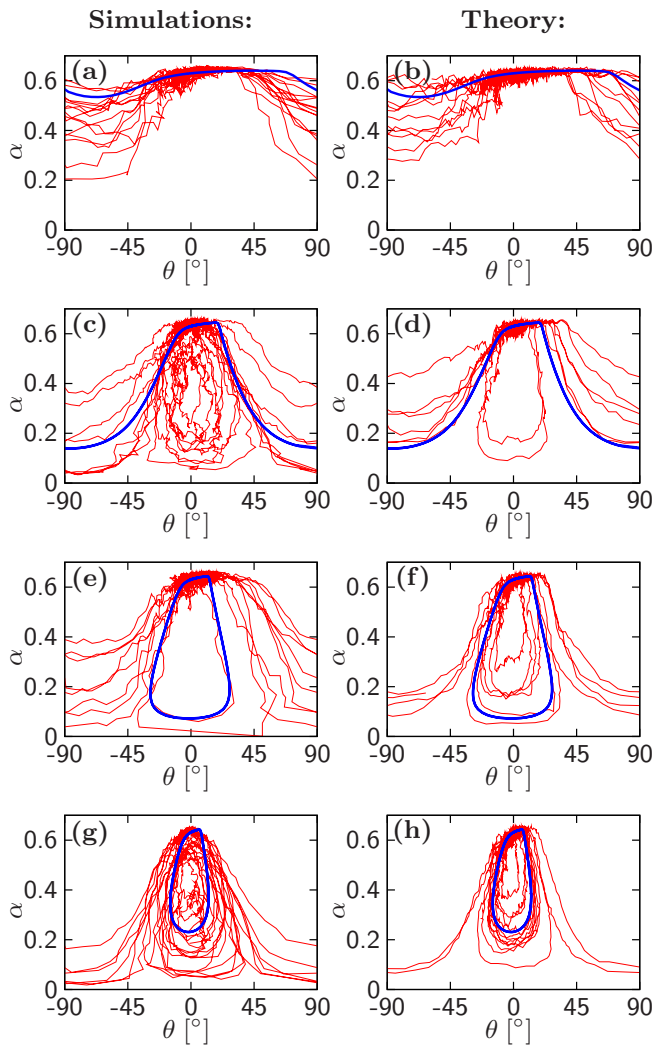}{
}
\caption{(Color online)
Trajectories in the $\theta$-$\alpha$ plane for $\lambda=4$, both with 
(thin red lines) and without 
(thick blue lines) thermal noise. The trajectories in (a), (c), (d), 
and (g) are 
obtained from simulations, whereas the curves in (b), (d), (f), and 
(h) are calculated from the generalized KS theory with noise and 
$b_1=1/(\lambda+1)$. The reduced shear rates are 
$\dot{\gamma}^*=5$
for (a) and (b), 
$\dot{\gamma}^*=17$
for (c) and (d), 
$\dot{\gamma}^*=22$
for (e) and (f), and
$\dot{\gamma}^*=38$
for (g) and (h).
In all plots, the corresponding theoretical trajectory
according the generalized Keller-Skalak theory without thermal noise is 
shown as a thick blue (dark) line.}
\label{fig:dynregsimtheo}
\end{figure}

\subsection{TB-SW Transition}
\label{sec:dynregimessimulations}

To investigate the TB-SW transition, we consider a fixed viscosity contrast 
of $\lambda=4$, and perform simulations for four different reduced shear 
rates $\dot{\gamma}^*=5, 17, 22$, and $38$.
The locations of these four shear rates in the dynamical phase
diagram are indicated in Fig.~\ref{fig:hirochisphasediagr}.
The resulting trajectories in the $\theta$-$\alpha$ plane are shown  
on the left-hand side of Fig.~\ref{fig:dynregsimtheo}.
In this representation, closed cycles indicate swinging events, whereas 
trajectories spanning the full $[-\pi/2,+\pi/2]$ range of $\theta$  
are tumbling events. Obviously, thermal noise has a large
impact on the vesicle dynamics. In particular, at small inclination angles 
$\theta \simeq 0$, small thermal fluctuations are decisive 
for the vesicle to perform a tumbling or swinging cycle.

The simulation data of Fig.~\ref{fig:dynregsimtheo} suggest that the 
TB-SW transition point is located between $\dot{\gamma}^*=17$ and $22$.
This is about a factor $2$ larger than the prediction 
$\dot{\gamma}^*=9$ of generalized KS theory.
In the generalized KS theory, a possible source of error can be found 
in the estimate of $b_0$ and $b_1$, which have both been calculated in 
the circular limit, see Eq.~(\ref{eq:b_factors}).
Therefore, we also calculate the phase diagram with 
$b_1$ reduced by a factor $2/3$, {\em i.e.} with $b_1=1/(\lambda+1)$, 
which gives a better agreement with our simulations for non-circular 
vesicles with $A^*=0.7$
(see Fig.~\ref{fig:hirochisphasediagr}). In this case,
the effect of thermal noise on trajectories is found to be very similar
as in the simulations
for all four reduced shear rates (see Fig.~\ref{fig:dynregsimtheo}).
Therefore, the deviations between the generalized KS theory and
simulations can be alleviated by a small modification of prefactors.
Further theoretical developments are needed to determine the  
prefactors $b_0$ and $b_1$ analytically for non-circular shapes.
We conclude that generalized KS theory provides a good description of 
vesicle dynamics in shear flow in both two and three spatial
dimensions.  

Elastic capsules~\cite{chan93,navo98} and red blood cells~\cite{Abkarian2007} 
can also exhibit a swinging motion. However, 
the angle $\theta(t)$ is always positive during these oscillations --- 
unlike SW of fluid vesicles. The physical mechanism is an energy 
barrier for the TT rotation caused by the membrane shear elasticity
and the anisotropic shape of the spectrin network 
\cite{Abkarian2007,skot07,kess08}.
Although, a vesicle in $2$D (a closed string) does not have membrane shear
elasticity, an energy barrier for the TT rotation can be introduced
by inhomogeneities in the spontaneous curvature \cite{sui07}.
In the future, it will be interesting to investigate the coupling of 
different swinging mechanisms in composite membranes.

\section{Lift Force} 
\label{sec:liftforces}

\begin{figure}
        \oneeps{width=8.6cm}{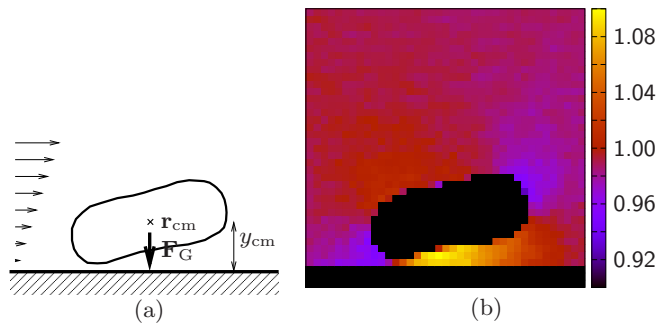}{
        }
\caption{(Color)
(a) Shape and (b) pressure field of a tank-treading vesicle  
under shear flow close to a wall, in steady state  
with viscosity contrast $\lambda=3$. 
The color code is expressed in units of $n_{\rm s} k_{\rm B}T$. 
The hydrodynamic lift force is balanced by an external 
gravitational force $F_{\rm G}=14k_{\rm B}T/a$, where the average 
distance from the wall is $y_{\rm {cm}}=7.96a$.
See also movie \cite{movie_S2}.
}
\label{fig:liftvessnapshots}
\end{figure}

We now consider a vesicle under the combined effect of a shear flow and
a gravitational force $F_{\rm G}$, see Fig.~\ref{fig:liftvessnapshots}(a). 
The vesicle moves towards or away from the wall until
gravitational $F_{\rm G}$ and lift forces $F_{\rm L}(y_{\rm {cm}})$ 
balance each other (see also movie \cite{movie_S2}). 
In this steady state, the lift force $F_{\rm L}(y_{\rm cm})$ equals 
the gravitational force in magnitude.

Fig.~\ref{fig:liftvessnapshots}(b) shows the pressure field in the outer 
fluid for the steady-state configuration of a tank-treading vesicle. 
The hydrodynamic lift force is the integral of the pressure forces over 
the membrane contour. The higher pressure in the gap between the vesicle 
and the wall is responsible for the lift force. 
Fig.~\ref{fig:liftvessnapshots}(b) also nicely demonstrates 
that there is a lower pressure at the two caps of 
the vesicle, which is the origin of vesicle elongation.
The hydrodynamic lift force is a pressure force which is of purely 
viscous nature -- in contrast to {\em e.g.} aerodynamic forces
acting on the wings of an airplane, which are caused by inertial forces.
 
\begin{figure}
        \oneeps{width=8.6cm}{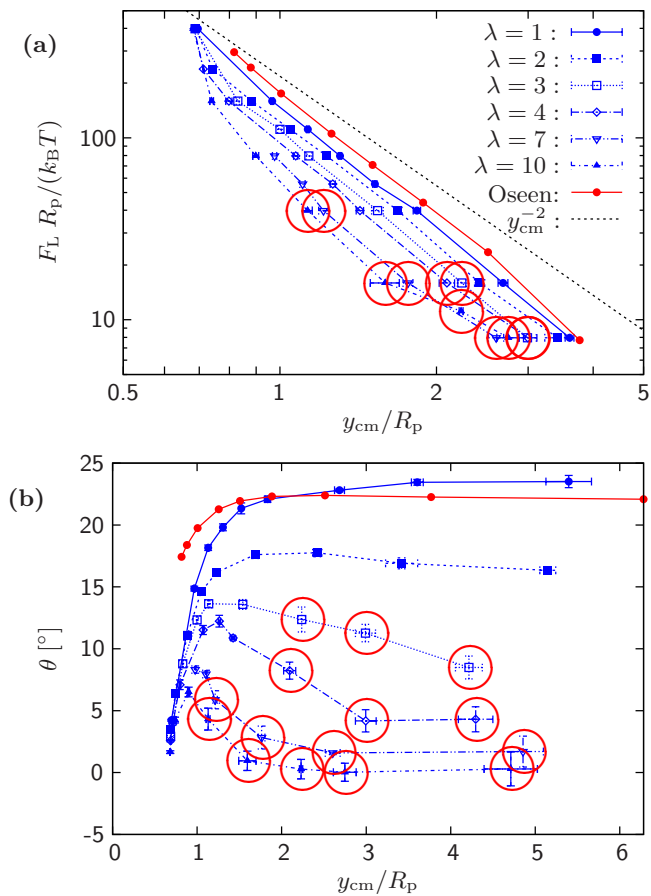}{
}
\caption{(Color online) (a) Lift forces and (b) average inclination angle 
as a function of the average wall distance $y_{\rm {cm}}$  
of the vesicle, from MPC simulations with $\dot{\gamma}^*=3.6$
and from boundary-integral (Oseen) calculations, as indicated.
The legend in (a) applies to both plots. Simulation points marked
by big (red) circles refer to tumbling vesicles.  
For comparison, a line with the power-law dependence 
$y^{-2}_{\rm {cm}}$ is plotted in (a).
The right-most data points in (b) correspond 
to $F_{\rm G}=0$, so that they cannot be shown in the double-logarithmic
presentation in (a). 
}
\label{fig:thetavsy}
\end{figure}

\begin{figure}[t]
   \oneeps{width=8.6cm}{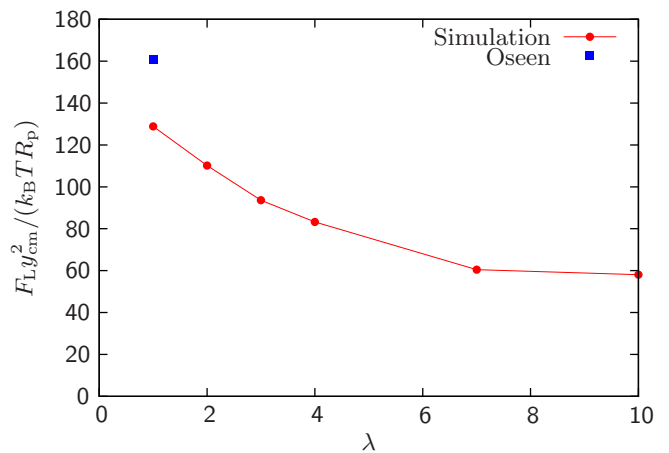}{
}
\caption{(Color online)
Amplitude of the lift force, $F_{\rm L} y_{\rm cm}^2/(k_{\rm B}T R_{\rm p})$, 
as a function of the viscosity contrast $\lambda$. The amplitudes 
are fits to the curves in Fig.~\ref{fig:thetavsy}(a) for which 
vesicles are not in direct contact with the wall.
}
\label{fig:FLamplitude}
\end{figure}

The dependence of the hydrodynamic lift force on the wall distance 
is shown in Fig.~\ref{fig:thetavsy}(a) --- calculated as 
$\langle y_{\rm cm} \rangle$ for fixed gravitational force in the 
simulations, and as lift force at fixed $y_{\rm cm}$ in the Oseen 
calculations, as explained in detail in Sec.~\ref{sec:memb_int} and 
Sec.~\ref{sec:oseen} above.
Lift forces of vesicles with $\lambda\le 4$ can be well described by a 
power-law $F_{\rm L}\propto y^{-2}_{\rm {cm}}$ for 
$F_{\rm L}\le 2.5k_{\rm B}T/R_{\rm p}$, corresponding to 
$y_{\rm {cm}} \gtrsim R_{\rm p}$.
For these distances, the vesicle is not in direct contact with the wall. 
At applied gravitational forces larger than $2.5k_{\rm B}T/R_{\rm p}$, 
the vesicle touches the wall ($y_{\rm {cm}} \lesssim R_{\rm p}$). 
However, the distance $y_{\rm {cm}}$ between the center of mass and the
wall can be reduced 
even further by vesicle deformation. The $1/y^2_{\rm {cm}}$ 
dependence does not apply in this regime. Finally, the constraints of 
fixed enclosed area $A$ and fixed perimeter $L_{\rm p}$ keep the  
wall distance larger than $y_{\rm {cm}}\gtrsim 0.628 R_{\rm p}$.

Fig.~\ref{fig:thetavsy}(a) shows that the lift forces decrease with 
increasing viscosity ratio $\lambda$ for a fixed wall distance $y_{\rm {cm}}$.
This behavior is analyzed in more detail in Fig.~\ref{fig:FLamplitude}, where 
the amplitude $F_{\rm L}y^2_{\rm {cm}}$ of the lift force is plotted 
as a function of the viscosity contrast $\lambda$.
Although solid colloidal particles of elliptical shape experience 
no net lift force~\cite{pozr06},
tumbling vesicles with finite $\lambda$ obtain lift force
due to an asymmetry of its shape deformations and a small tank-treading
component (compare Fig.~\ref{fig:vessnapshots}(c)).

For vesicles in three dimensions, both boundary-integral simulations
\cite{Sukumaran2001} as well as theoretical studies
\cite{Olla1997,Olla1997a} show a $1/y^2_{\rm {cm}}$ dependence of 
the lift force for vesicles far from the wall. The theory of Olla 
\cite{Olla1997,Olla1997a} assumes an ellipsoidal shape for the vesicles
with half axes $a_1,a_2,a_3\ll y_{\rm {cm}}$. It is not possible 
in this case to derive expressions for two dimensions by taking the
limit $a_3\to\infty$ --- as done in App.~\ref{sec:KS_derivation} for  
KS theory of vesicles in unbounded flows --- because this limit is
inconsistent with the assumption $a_1,a_2,a_3\ll y_{\rm {cm}}$.
Therefore, instead of an analytical theory, we perform boundary-integral 
calculations of $2$D elliptical vesicles with $\lambda=1$ 
in the presence of a wall, as described in Sec.~\ref{sec:oseen}.
For the results in Fig.~\ref{fig:thetavsy}(a), the effect of the opposite wall
at  $L_y=10 R_{\rm p}$ is also taken into account by plotting 
$F_{\rm L}(y_{\rm {cm}})-F_{\rm L}(L_y-y_{\rm {cm}})$, where 
$F_{\rm L}(y_{\rm {cm}})$ and $F_{\rm L}(L_y-y_{\rm {cm}})$ are obtained 
from two independent boundary-integral calculations. 
Of course, a more precise calculation would require the use the two-wall 
Oseen tensor \cite{Pozrikidis1992} instead of the half-space Greens 
function. However, as long as the distance $L_y$ between the 
two walls is sufficiently large, 
$F_{\rm L}(y_{\rm {cm}})-F_{\rm L}(L_y-y_{\rm {cm}})$ 
is a good approximation of the two-wall lift force. 
Fig.~\ref{fig:thetavsy}(a) as well as Fig.~\ref{fig:FLamplitude} show 
that the boundary-integral calculation indeed agrees nicely with the 
corresponding MPC simulation 
for $\lambda=1$. The amplitudes $F_{\rm L}{y_{\rm {cm}}}^2$
only differ by about 25\%. Reasons for this deviation are that the 
boundary-integral calculation is done for elliptical shapes, whereas 
vesicles in simulations are closer to the equilibrium shape (compare 
Fig.~\ref{fig:oseen_drawing} and Fig.~\ref{fig:liftvessnapshots}).
Moreover thermal fluctuations in the MPC simulations may cause differences.

\begin{figure}[t]
   \oneeps{width=9.0cm}{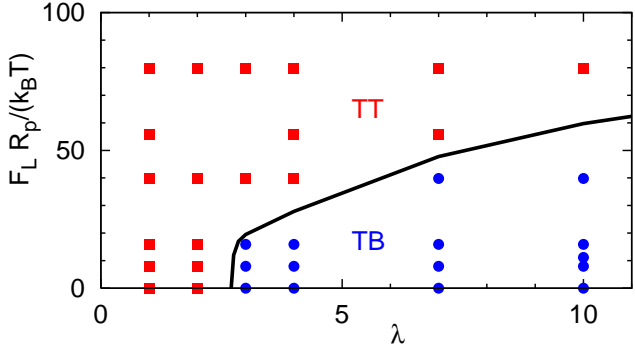}{
}
\caption{(Color online)
Dynamic phase diagram of tank-treading and tumbling states as a function
of the gravitational force $F_{\rm G}$ and the viscosity
contrast $\lambda$. The reduced shear rate is $\dot\gamma^*=3.6$. 
Circles ($\bullet$) indicate tumbling motion (TB), squares ($\blacksquare$) 
tank-treading motion (TT). The line for the phase boundary is a guide 
to the eye. 
}
\label{fig:lift_phasediag}
\end{figure}
      
Fig.~\ref{fig:thetavsy}(a) shows that tumbling is suppressed when the 
gravitational force exceeds a threshold value, depending on the viscosity 
contrast $\lambda$. In order to perform a 
tumbling motion, the center-of-mass distance $y_{\rm {cm}}$ has to be
on the order of or larger than the long vesicle axis $a_1$.
However, for larger gravitational 
forces, the center-of-mass distance $y_{\rm {cm}}$ becomes smaller than 
$a_1$, such that even vesicles with high viscosity
contrasts do not tumble. Even if $y_{\rm {cm}}$ is slightly larger than 
$a_1$, the vesicle cap has to come so close to the wall that the resulting 
pressure forces prevent the inclination angle to reach $\pi/2$.
This leads to the dynamical phase diagram shown in 
Fig.~\ref{fig:lift_phasediag}. When the gravitational force is large,
tumbling is suppressed and the vesicle displays a tank-treading motion
at the wall. With increasing $\lambda$, the gravitational force 
necessary to prevent tumbling increases.

The dependence of the inclination angle $\theta$ on the wall 
distance $y_{\rm {cm}}$ is shown in Fig.~\ref{fig:thetavsy}(b). 
Without a gravitational force, the 
lift force caused by the upper wall at $y=10R_{\rm p}$ compensates 
the lift force of the lower wall when the vesicle is in the center, at
$y_{\rm {cm}}=5R_{\rm p}$. Since the lift forces are
very small nearby, strong fluctuations are 
observed in the wall distances for small $F_{\rm G}$.

As long as a vesicle is tank-treading, its inclination angle $\theta$
decreases when it approaches the wall. 
Even if the vesicle does not touch the wall, the pressure at the lowest
part of the membrane is highest (see Fig.~\ref{fig:liftvessnapshots}(b)) 
such that it causes a torque which lowers $\theta$. For very small 
wall distances, the vesicle comes into direct contact with the wall, 
where the repulsive wall potential causes an additional 
torque, which decreases $\theta$ even further, until the vesicle is finally 
completely parallel to the wall.
    
Vesicles with $\lambda\ge 3$ start to tumble at sufficiently large 
wall distances.  Since vesicles with viscosity ratios $\lambda=3$ 
and $\lambda=4$ are still tank-treading most of the time and only 
occasionally perform a tumbling motion, their inclination angles are
non-zero, whereas for $\lambda=10$ and $F_{\rm G}\le 1$, the average 
inclination angle $\theta$ essentially vanishes (see 
Fig.~\ref{fig:thetavsy}(b)).
  
\begin{figure}
   \oneeps{width=8.6cm}{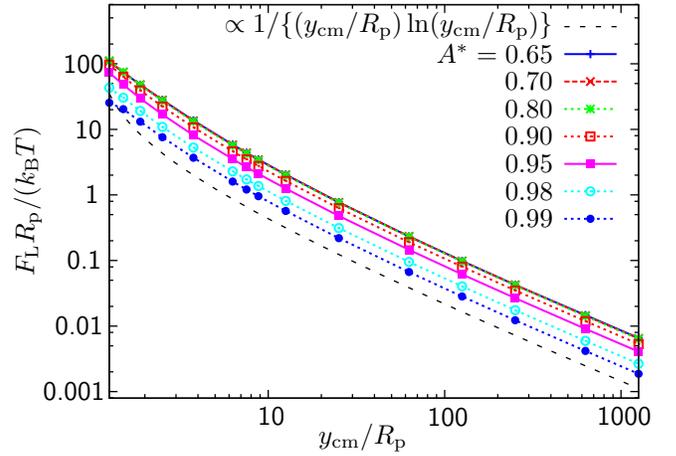}{
}
\caption{(Color online)
Hydrodynamic lift force $F_{\rm L}$ on vesicles with varying $A^*$ vs. 
wall distance $y_{\rm {cm}}$ obtained from boundary-integral calculations.
}
\label{farfield}
\end{figure}

\begin{figure}
   \oneeps{width=8.4cm}{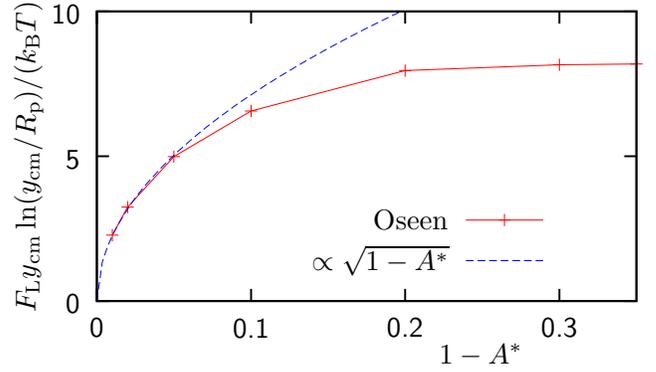}{
}
\caption{(Color online)
Amplitude of the lift force $F_{\rm L}$ as a function of $1-A^*$. The 
data points are fits to the curves in Fig.~\ref{farfield} for 
$y_{\rm {cm}}\ge 60R_{\rm p}$. These data points (Oseen) are fitted to 
a $\sqrt{1-A^*}$ dependence for small values of $1-A^*$.
}
\label{amplitudesVSRedAreas}
\end{figure}

We employ the boundary-integral approach to calculate the dependence of
the hydrodynamic lift force of vesicles with $\lambda=1$ on the 
reduced area $A^*$ in the absence of thermal fluctuations.
Also, since this method is not 
restricted by the system size (as simulations), the lift forces can be
calculated even for very large wall distances. Fig.~\ref{farfield}
shows the hydrodynamic
lift force as a function of the wall distance $y_{\rm {cm}}$  
for different reduced areas $A^*$ in the range $0.65\le A^*\le 0.99$. 
This plot shows that the
$y_{\rm {cm}}^{-2}$ dependence of the hydrodynamic lift force
$F_{\rm L}$ only holds for distances $y_{\rm {cm}}\lesssim 5 R_{\rm p}$ 
from the wall. For larger distances $y_{\rm {cm}}$, a crossover to a 
$1/\left[(y_{\rm {cm}}/R_{\rm p})\ln \left(y_{\rm {cm}}/R_{\rm p}\right)\right]$ 
dependence 
is obtained. This power law with a logarithmic correction  
fits perfectly the numerical data of the boundary-integral calculation for 
$y_{\rm cm}\gtrsim 2R_{\rm p}$ for all considered reduced areas. 
Fig.~\ref{amplitudesVSRedAreas} shows
the amplitudes $K=F_{\rm L} y_{\rm {cm}}\ln(y_{\rm {cm}}/R_{\rm p})$
of the lift forces $F_{\rm L}$ vs. $1-A^*$
in the far-field limit. These amplitudes $K$ are determined by
fitting the expression 
$F_{\rm L}=K /\left[(y_{\rm {cm}}/R_{\rm p})\ln \left(y_{\rm {cm}}/R_{\rm p}\right)\right]$ 
to all data with $y_{\rm {cm}}\ge 60R_{\rm p}$.  
For $1-A^*=0$, {\em i.e.} for a circular vesicle, the lift force vanishes, 
which directly follows from the time-inversion symmetry of the Stokes equation. 
For small deviations from the circular shape, the lift force rapidly 
increases with $1-A^*$, and follows a power-law dependence 
$K\sim \sqrt{1-A^*}$ $\simeq \sqrt{\Delta_{\rm p}/\pi}$,  with the excess 
length $\Delta_{\rm p}=L_{\rm p}/R_{\rm A} - 2\pi$, for $1-A^*\lesssim 0.1$ 
(see Fig.~\ref{amplitudesVSRedAreas}).
We observe a monotonic increase of the 
lift force with increasing $1-A^*$ over the whole range of 
reduced areas $A^*$. 
 
Since boundary-integral calculations do not take into account
thermal fluctuations, lift forces can be determined  
for large $y_{\rm {cm}}$ (only limited by numerical accuracy). 
However in simulations as well as in real systems, lift forces at 
large $y_{\rm {cm}}$ have a vanishing effect compared to thermal noise. 
Moreover, with very large wall distance 
$y_{\rm {cm}}\sim \sqrt{\eta_{\rm {out}}/\rho\dot{\gamma}}$, 
inertial effects are not negligible, so that the Stokes approximation 
becomes less reliable \cite{Olla1997a}.

\section{Summary and Conclusion}

We have studied the dynamics of vesicles in shear flow in a 
two-dimensional model system.
This system shows a variety of interesting dynamical phenomena. 

First, we have investigated  
the effect of the viscosity contrast $\lambda$, {\em i.e.} the ratio between 
the inner and outer viscosities of a vesicle, on the dynamics
in unbounded flows.  With increasing $\lambda$,  
the sequence from ``tank-treading'' over ``swinging'' to 
``tumbling'' motion is generically observed --- except for small
shear rates $\dot\gamma$, where the intermediate swinging phase is
absent.
Thus, the swinging phase appears in the phase diagram of $2$D vesicles 
under shear in the same way as it was found previously for $3$D 
ellipsoidal vesicles. However, the mechanism of swinging is 
different in two and three dimensions. While in $3$D, ellipsoidal 
deformations are sufficient to obtain swinging, in $2$D higher-order 
undulation modes are required. Thermal fluctuations play an important
role; they lead to a smooth crossover between the dynamical states,
with intermittent tumbling and tank-treading motions. 
Our simulations are in semi-quantitative agreement 
with a theoretical description
based on a generalized Keller-Skalak approach.

Second, we have investigated the behavior of vesicles near walls.
Close to a wall, tumbling is strongly suppressed. Furthermore, the vesicle 
is repelled from the wall by the hydrodynamical
lift force.  We have found by boundary-integral calculations that the 
hydrodynamic lift force decays with increasing
wall distance $y_{\rm {cm}}$ like $1/\left(y_{\rm {cm}}\ln y_{\rm {cm}}\right)$.
However, for small wall distances -- in particular in the regime of 
the MPC simulations --
an effective $y_{\rm {cm}}^{-2}$ dependence is observed. With increasing 
viscosity contrast, the lift force becomes weaker, as the vesicle becomes 
less deformable. The lift force also decreases with increasing 
reduced area $A^*$, and vanishes in the circular limit. We find that
our numerical data are well described by a $\sqrt{1-A^*}$ dependence.

Our results show that there is a different behavior of the lift force
at intermediate and large distances from the wall, and that the 
lift force decreases significantly with increasing viscosity contrast. 
This may shed some light on the behavior  
in three dimensions, where experiments show a dependence of the lift 
force on the wall-membrane distance $h$, which decays as $h^{-1}$ for 
distances smaller than the vesicle radius \cite{Abkarian2002}, whereas
a $y_{\rm cm}^{-2}$ decay has been found theoretically in a small
range $1.1 \lesssim y_{\rm cm}/R_{\rm p} \lesssim 1.25$ of wall
distances \cite{Sukumaran2001}. Thus, we hope that our results 
will stimulate new experiments and simulations in $3$D over a wider
range of wall distances, reduced volumes, and viscosity contrasts.

\begin{acknowledgments}
Sebastian Messlinger acknowledges a fellowship of the International 
Helmholtz Research School ``BioSoft". Benjamin Schmidt thanks the 
DAAD for financial support through the 
RISE (Research Internships in Science and Engineering) program and 
for giving him the opportunity of a visit at the Research Center J\"ulich. 
\end{acknowledgments}

\begin{appendix}
\section{Derivation of 2D Generalized Keller-Skalak Theory}
\label{sec:KS_derivation}

\subsection{Keller-Skalak Theory in Two Dimensions}

Keller and Skalak \cite{KELLER1982} derived analytical expressions for
the inclination angle $\theta$ and the average angular velocity $\omega$ 
for $3$D vesicles of fixed ellipsoidal shape, 
with $(x/a_1)^2 + (y/a_2)^2 + (z/a_3)^2=1$, based on the Jeffery 
theory \cite{jeff22}.

Although the KS theory is formulated for vesicles in three dimensions, 
it is straightforward to transfer it to two-dimensional systems by 
simply taking the limit $a_3 \to\infty$.
The resulting cylindrical three-dimensional geometry is equivalent 
to a $2$D vesicle with the shape of an ellipse, $(x/a_1)^2 + (y/a_2)^2=1$
with $a_1 \ge a_2$.
Let $S'$ be the frame which has its origin at the center of the ellipse, 
and the $x'$ direction points into the direction of the long axis. Then 
the local velocity $\vec{v}'$ of an element of the tank-treading 
membrane is assumed to be
\begin{equation}
    (v_x',v_y')   =   \omega (- \frac{a_1}{a_2}x_2',\frac{a_2}{a_1}x_1').
\end{equation}
in the frame $S'$. We define the auxiliary variables
\begin{equation*}
    f_0 := \frac{1-{\alpha_{\rm D}}^2}{1+{\alpha_{\rm D}}^2},  \qquad 
    f_1 := \frac{1-{\alpha_{\rm D}}^2}{8\alpha_{\rm D}},  \qquad 
    f_2 := \frac{1+{\alpha_{\rm D}}^2}{2},
\end{equation*}
where $\alpha_{\rm D}= (a_1+a_2)/(a_1-a_2)$.
The balance of torques on the membrane and the assumption that the work 
done on the vesicles by the shear flow is dissipated in the interior of 
the vesicle leads to the non-linear differential equation 
\begin{eqnarray}
  \frac{d\theta}{dt} & = &
       \frac{\dot{\gamma}}{2}\left[-1+B(\alpha_{\rm D},\lambda)
			           \cos(2\theta)\right]
      \label{equ:KSequation}\\
  B(\alpha_{\rm D},\lambda) & = &  f_0\left\{f_1+ \frac{f_1^{-1}}
         {1+f_2(\lambda-1) }\right\} .
\end{eqnarray}
Furthermore, the average angular velocity $\omega$ is found to be
\begin{equation}
   \frac{\omega}{\dot{\gamma}}= \frac{\cos(2\theta)}
        {2f_1\{1+f_2(\lambda-1)\}}
\end{equation}

\begin{figure}
    \oneeps{width=8.6cm}{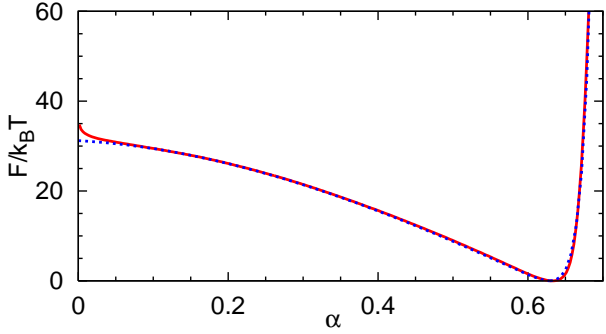}{
}
\caption{ \label{fig:fcv}
(Color online)
Free energy $F(\alpha)$ 
of the two-dimensional vesicle relative to
the prolate shape
for $A^* = 0.7$, $\kappa/l=50 k_{\rm B}T$, 
$k_{\rm {sp}}=10^4 k_{\rm B}T/a^2$, and $k_{\rm A}=80 k_{\rm B}T/a^4$.
The solid (red) and dashed (blue) lines represent 
the simulation data and fitted curve, respectively.
}
\end{figure}

\subsection{Shape Equation in Two Dimensions}

The vesicle shape is expanded in Fourier modes with polar angle $\phi$ as 
${\bf r}(\phi)=R_{\rm A}{\bf e}_r\{1+\sum u_m \exp(im\phi/\sqrt{2\pi})\}$.
Based on the Stokes approximation and perturbation theory,
the dynamics of a quasi-circular vesicle is described by~\cite{fink08}
\begin{equation}
\label{eq:lamb}
   \frac{\partial u_{m}}{\partial t} = \frac{i \dot\gamma m}{2} u_{m}- 
         \frac{\kappa\Gamma_m E_m}{\eta_{\rm {out}} R_{\rm A}^3}{u_{m}}
            \mp i h\dot\gamma \delta_{m,\pm 2},
\end{equation}
where $\Gamma_m=|m|/2(\lambda+1)(m^2-1)$, $E_m=(m^2-1)(m^2-3/2+\sigma)$, and
$h=\sqrt{2\pi}/2(\lambda+1)$. A Lagrange multiplier $\sigma$ keeps the
perimeter $L_{\rm p}$ constant. Following the procedure for $3$D~\cite{Noguchi2007a},
we decompose $u_{\pm 2}$ into amplitude and phase, 
$u_{\pm 2}=r \exp(\mp 2i \theta)$, and replace the force $2 \kappa E_m r$  
by $\partial F/\partial r$. Then, Eq.~(\ref{equ:alphaevolution}) is 
obtained with $\alpha= 3r/\sqrt{2\pi}+ O(r^2)$.

The free energy $F(\alpha)$ for the same simulation parameters
calculated with a version of the 
generalized-ensemble Monte Carlo method \cite{nog05} (see Fig. \ref{fig:fcv}):
The vesicle conformations are sampled under the uniform distribution of $\alpha$,
and then the canonical distribution is obtained by the reweighing.
Instead of an interpolation~\cite{nog04,nog05,Noguchi2007a},
we use fit functions here to obtain smooth functions.
The force is fitted as a function
$-(1/k_{\rm B}T)\partial F/\partial \alpha= 9+180 \alpha -110 \alpha^2 -\exp(80 \alpha -46)$.
We obtained the relation $\alpha_{\rm D}= 2\alpha/3 +0.14 \alpha^4$
by fitting for the ellipse of $A^*=0.7$.

\begin{figure}
    \oneeps{width=5cm}{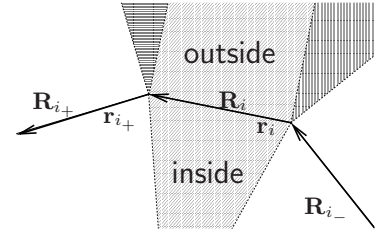}{
}
\caption{
Spatial regions for the membrane collision between MPC particles
and membrane monomers. MPC particles 
located in the dark-gray regions 
perform a two-body collision with the monomers $i$ or $i_+$.
MPC particles in the light-gray regions collide with both
monomers.  For further explanations, see text.
}
\label{fig:bondcollrule}
\end{figure}

\section{Membrane Collisions with the Solvent}
\label{sec:membrane_coll}

For the modeling of an impermeable membrane, interior and exterior
solvent particles have to stay on the appropriate side of the 
membrane.
Depending on the position of the MPC particle with
respect to membrane bonds, either one or two membrane monomers 
participate in a membrane collision. The new velocities of the
$n_{\rm {coll}}$ particles (the MPC particle
and the $n_{\rm {coll}}-1$ membrane monomers) after the collision are then
\begin{equation} \label{equ:membranecollision}
  \vec{v}_{i,{\rm {new}}}=2\left(\vec{v}_{\rm {cm}} 
       + \vec{\omega}\times\vec{r}_{i,c}\right) - \vec{v}_i,
\end{equation}
where $v_{\rm {cm}}$ is the center-of-mass velocity of the 
$n_{\rm {coll}}$-body system, $\omega$ its angular velocity, and  
$\vec{r}_{i,c}$ are the particle positions relative to the 
center of mass of the $n_{\rm {coll}}$-body system.   
This is a bounce-back collision for the relative velocities which
conserves both the total translational and angular momenta.
The spatial regions for the selection of colliding MPC particles and 
membrane monomers are illustrated in Fig.~\ref{fig:bondcollrule}.

\end{appendix}

\bibliographystyle{apsrev}

\end{document}